\documentclass[traditabstract]{aa} 
\usepackage{txfonts,graphicx}

\begin{document}

\title{Variegate galaxy cluster gas content: Mean fraction,
scatter, selection effects, and covariance with X-ray luminosity.
}
\titlerunning{Variegate galaxy cluster gas content} 
\author{S. Andreon\inst{1} \and J. Wang\inst{2} \and G. Trinchieri\inst{1} \and A. Moretti\inst{1} 
\and A. L. Serra\inst{3}
}

\authorrunning{Andreon et al.}
\institute{
$^1$INAF--Osservatorio Astronomico di Brera, via Brera 28, 20121, Milano, Italy; \email{stefano.andreon@brera.inaf.it}\\
$^2$Dep. of Physics and Astronomy, University of the Western Cape, Cape Town 7535, South Africa\\
$^3$Dip. di Fisica, Universit\`a degli Studi di Milano, via Celoria 16, 20133, Milano, Italy\\
}
\date{Accepted ... Received ...}
\abstract{
We use a cluster sample selected independently of the intracluster
medium content with reliable masses
to measure the mean gas mass fraction and its scatter, 
the biases of the X-ray selection
on gas mass fraction, and the covariance between the X-ray luminosity and
gas mass. The sample is formed by
34 galaxy clusters in the nearby ($0.050<z<0.135$) Universe,
mostly with $14<\log M_{500}/M_\odot \lesssim 14.5$, and with 
masses calculated with the caustic technique.
First, we found that integrated gas density profiles have similar shapes,
extending earlier results based on subpopulations of clusters
such as those that are relaxed or X-ray bright for their mass. Second, the X-ray unbiased
selection of our sample allows us to unveil a variegate population of clusters; 
the gas mass fraction shows a scatter of $0.17\pm0.04$ dex,
possibly indicating a quite variable amount of feedback from cluster to cluster,
which is larger than is found in previous samples
targeting subpopulations of galaxy clusters, such as relaxed or X-ray
bright clusters. The similarity of the gas density profiles induces an
almost scatterless relation between X-ray luminosity, gas mass, and
halo mass, and modulates selection effects in the halo
gas mass fraction: gas-rich clusters are preferentially included in
X-ray selected samples. The almost
scatterless relation also fixes the relative scatters and slopes 
of the $L_X-M$ and $M_{gas}-M$ relations and makes core-excised
X-ray luminosities and gas masses fully covariant. Therefore, 
cosmological or astrophysical
studies involving X-ray or SZ selected samples need to account for both selection
effects and covariance of the studied quantities with X-ray luminosity/SZ strength.  
}
\maketitle

\section{Introduction}

Scaling relations between X-ray luminosity, gas mass, and halo mass are useful
both for cosmological studies and to learn about the physical processes that
shape the intracluster medium. Their observational derivation is however
hampered by selection effects. All methods used to select galaxy clusters via one of their
constituting and observable parts, such as  galaxies, 
intracluster medium, or dark matter, 
are subject to biases due to scatter between the mass and observable (or selecting quantity), or
its dependency on other physical properties of the clusters.
Even the direct observation of dark matter, via weak lensing,
is subject to scatter with mass due to cluster triaxiality, 
large-scale structure and
intrinsic alignments (Meneghetti et al. 2010; Becker \& Kravtsov 2011). 
As recent discussions in the literature show, there is increasing evidence
that X-ray surveys preferentially
select overluminous clusters with peaked central emission (Pacaud et al. 2007;
Andreon \& Moretti 2011; Andreon, Trinchieri \& Pizzolato 2011; Andreon \& Hurn 2013, 
Planck Collaboration IX 2011). 
The amount of bias depends on the scatter between the observable and
mass. A  
small scatter is often benign and has no significant consequences. 
A large scatter implies that 
typical (average) objects are under-represented, or even missing altogether, in
surveys. In the latter case, the bias
correction is hard at best because one needs to recover the width of a distribution
(the scatter) from a tail, which might not even be recognized as
such.  Therefore, knowledge of the
scatter is key to correcting  the bias, which in turn  is 
possible only if the sample is not censored too much. Biases are also possible when
dealing with samples complete in the selecting quantity, as illustrated by Fig.~5
in Giles et al. (2017) for an $L_X$-complete sample. This occurs because the sample
is heavily censored in the observable-mass plane.

A way to limit the bad effects of selection is to select the sample independently
of the quantity under study. After X-ray observations of single high-redshift clusters
(Andreon et al. 2008, 2009, 2011), an initial
low-redshift sample lacking reliable mass estimates (Andreon \& Moretti 2011),
in Andreon et al. (2016, paper I) we built (and observed in X-ray) a sample of
clusters selected independently
of the quantity under study and with reliable masses,
dubbed X-ray Unbiased Cluster Survey (XUCS). This sample allows us to derive
in the present work the mean gas mass fraction and its scatter
free of the complications arising from the selection function and 
without making any hypothesis on the behavior of the
unobserved population. This sample may provide a hint about why
recent works addressing the gas mass fraction 
and neglecting selection function effects found opposite results. Indeed,
Landry et al. (2013) found that clusters are richer in gas 
than found in Sun et al. (2009),
while Eckert et al. (2016) found the contrary.

Second, the availability of an X-ray unbiased sample  allows us to bring to light
the relative scaling of X-ray and gas mass with halo mass and the covariance
between them without the complications arising from the accounting
of the selection effects. Knowledge of the covariance is paramount to propagate
selection effects from the quantity used to select the sample (often
X-ray flux or luminosity) to the quantity of interest (i.e., gas mass).

As a spinoff, the comparison of the gas mass profiles, which is needed to derive
gas masses, may offers a hint about the reason for large variety of core-excised X-ray 
luminosities at a given mass found in Andreon et al. (2016, hereafter
Paper I). In that paper we found an intrinsic scatter of 0.5 dex,
much larger than the value (0.07 dex) inferred in the representative XMM-Newton 
cluster structure survey  (REXCESS; Pratt et al. 2009) 
and also larger than observed in a Planck-selected
sample of clusters of high mass (Planck Collaboration IX 2011). 
A large variety of core-excised X-ray 
luminosities at a given mass impact negatively on X-ray selected 
samples because it 
makes the survey selection function, $p(L_X,det|M,z)$, poorly known and hard to recover from
the bright tail of the population 
that enters in the sample.
The measurement of the gas mass profiles may allow us to discriminate 
the origin of the $L_X$ scatter, owing to variations from cluster to
cluster of the total gas content 
or to different shapes of gas density profiles, indicating 
a cluster-to-cluster variation in the  
redistribution of the gas within 
$r_{500}$\footnote{The radius $r_\Delta$ is the
radius within which the enclosed average mass density is $\Delta$
times the critical density at the cluster redshift.}.

Throughout this paper, we assume $\Omega_M=0.3$, $\Omega_\Lambda=0.7$, 
and $H_0=70$ km s$^{-1}$ Mpc$^{-1}$. 
Results of stochastic computations are given
in the form $x\pm y$, where $x$ and $y$ are 
the posterior mean and standard deviation. The latter also
corresponds to 68\% intervals because we only summarize
posteriors close to Gaussian in this way. Logarithms are in base 10.

\begin{figure}
\centerline{
\includegraphics[width=7truecm]{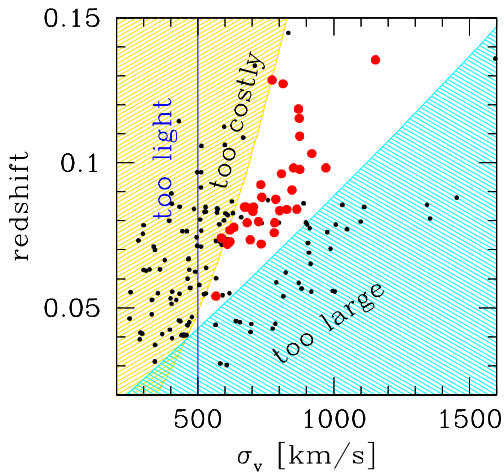}
}
\caption[h]{Sample selection of XUCS. The shadedless wedge is 
the observational optimal locus (clusters fit inside the
Swift XRT field of view yet are not too far, so they are not too
expensive in terms of exposure time). 
Points are high-quality clusters
in the C4 catalog, red points are those that compose the 
XUCS sample. 
}
\end{figure}

\section{Sample selection, cluster masses, and X-ray data}

Sample selection, halo mass derivation, and X-ray data are presented and 
discussed in paper I, to which we refer for details. 
In short, our XUCS sample
consists of 34 clusters in the very nearby universe 
($0.050<z<0.135$) extracted from the C4 catalog (Miller et al. 2005)
in regions of low Galactic absorption. The C4 catalog   
identifies clusters as overdensities in the local Universe
in a seven-dimensional space (position, redshift, and
colors; see Miller et al. 2005 for details) using
Sloan Digital Sky Survey data (Abazajian et al. 2004). As illustrated in Fig.~1, among all C4
high-quality clusters ($>30$ spectroscopic members within a radius of 1.5 Mpc, 55 on average) 
with a dynamical mass $\log M>14.2$ M$_\odot$ (derived from 
the C4 $\sigma_v>500$ km/s), to minimize exposure times
we selected the nearest which are also
smaller than the Swift XRT field of view
($r_{500}\lesssim 9$ arcmin).
Very few clusters (black points inside the shadedless wedge in Fig.~1) 
were later discarded because their centers fall too close to the boundary of the
XRT field of view (or even outside it) as a result of approximations in
the coordinates listed in the C4 catalog, or, for two clusters, because they are part of 
same dynamical entity, making caustic mass estimation unreliable (see Paper I for
more details). This leaves a final sample
of 34 clusters.

There is no X-ray selection in our sample, meaning that 1) the probability of
inclusion of the cluster in the sample is independent of its X-ray luminosity
(or count rate) and 2) no cluster is kept or removed on the basis of its
X-ray properties. In particular,
we do not select preferentially relaxed clusters, 
which probably represent  
a more  homogenous class of objects  than the whole
cluster population.
 
We collected the few X-ray observations present 
in the XMM-Newton or Chandra archives
and observed the remaining clusters with Swift 
(individual exposure times 
between 9 and 31 ks), as detailed in Table 1 of Paper I. Swift observations have 
the advantage of a low X-ray background (Moretti et al. 2009), 
making it extremely useful for sampling a cluster
population that includes low surface brightness clusters
(Andreon \& Moretti 2011).  

Caustic masses within $r_{200}$, $M_{200}$, were derived  following Diaferio \& Geller
(1997), Diaferio (1999), and Serra et al. (2011) and then converted into $r_{500}$ and $M_{500}$
assuming a Navarro, Frenk \& White (1997) profile with concentration $c=5$. Adopting
$c=3$ would change mass estimates by a negligible amount; see Paper I. 
Basically, the caustic technique performs a
measurement of the line-of-sight escape velocity and has the advantage of not assuming virial
equilibrium, which is assumed instead when estimating velocity-dispersion-based masses. It only uses
redshift and position of the galaxies to identify the caustics on the redshift diagram
(clustrocentric distance versus line-of-sight velocity), whose amplitude is a measure of the
escape velocity of the cluster. The
median number of members within the caustics is $116$ and the
interquartile range is $45$. The median mass of the
cluster sample is $\log M_{500} / M_\odot$ is $14.2$ and the interquartile range is
0.4 dex. The average mass error is 0.14 dex.

\begin{figure*}
\centerline{
\includegraphics[width=6truecm]{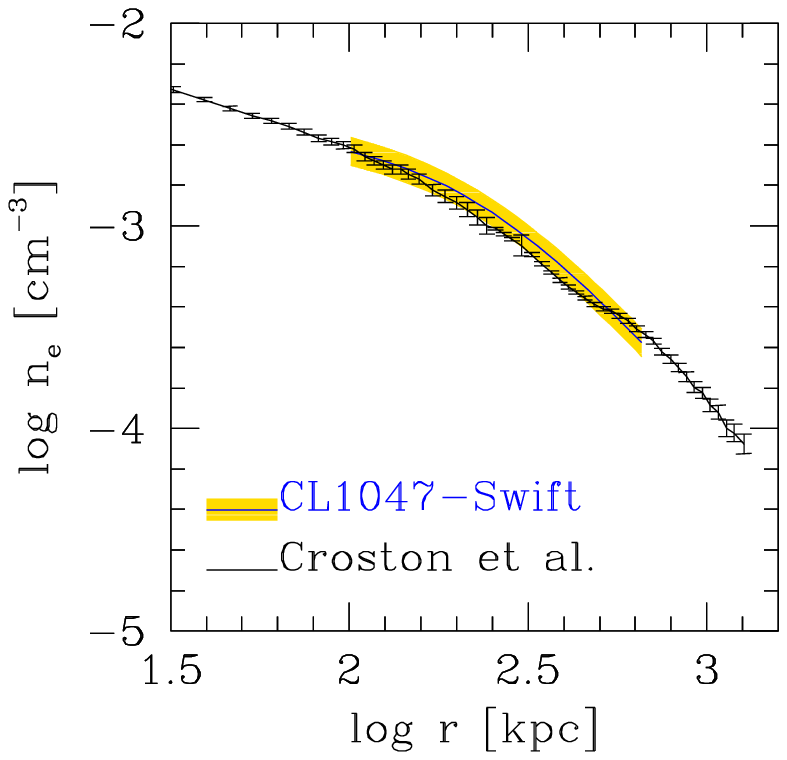}
\includegraphics[width=6truecm]{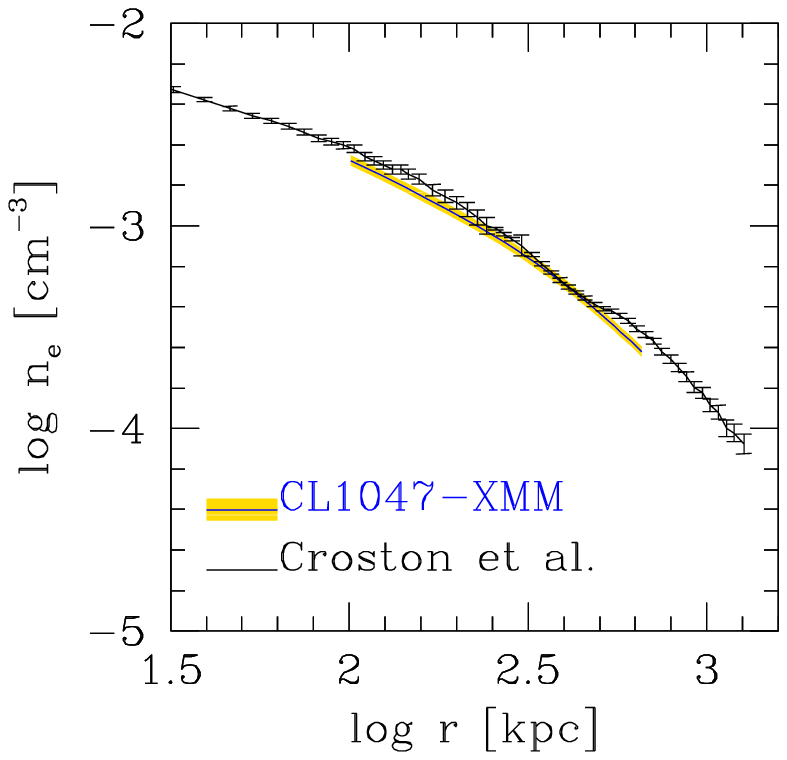}
\includegraphics[width=6truecm]{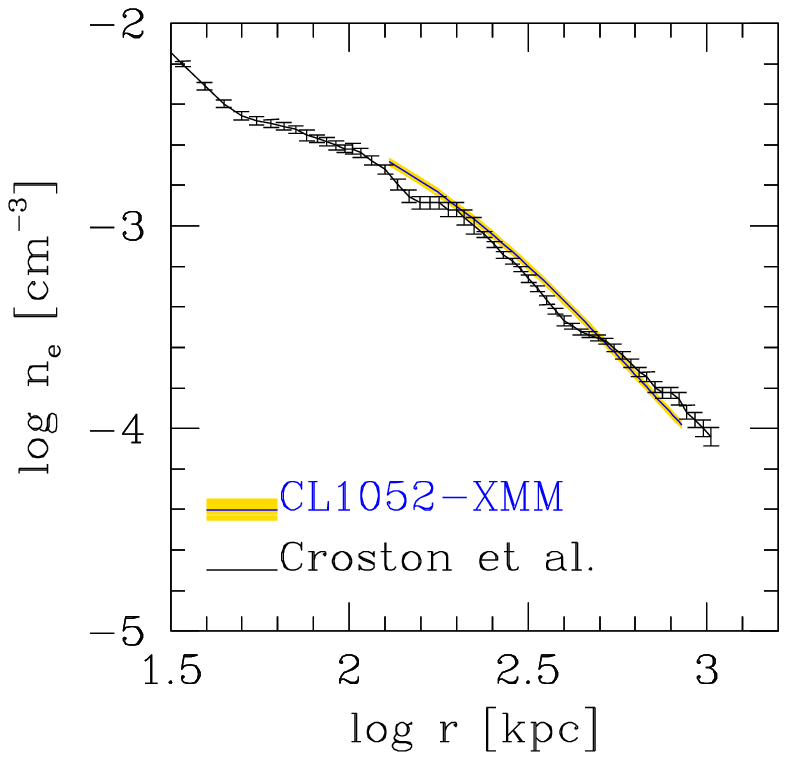}
}
\caption[h]{Deprojected electron density profile as derived by us (smooth solid 
line with shaded 68\% bounds)
and by Croston et al. (2008; points with error bars). 
The left-hand panel compares profiles derived from different
satellites, whereas the other two panels show analyses sharing (much the same) XMM photons. 
Profile smoothness and errors depend on assumptions; see text.
}
\end{figure*}

\section{X-ray analysis}

We used X-ray data to measure the surface brightness profile and  
the spectrum, whose normalization is needed to convert
the surface brightness profile into an emissivity profile. 

Data reduction was described in Paper I. Briefly,
we reduced the X-ray data using the standard data reduction procedures 
(Moretti et al. 2009; XMMSAS\footnote{http://xmm.esac.esa.int/sas} or
CIAO\footnote{http://cxc.harvard.edu/ciao/}). Point sources were detected by 
a wave detection algorithm and we masked pixels contaminated by them
when calculating spectral normalization and radial profiles.
In our analysis, we used exposure maps to calculate the effective
exposure time accounting for dithering, vignetting, CCD defects, gaps, and
excised regions. 

Since Swift observations are taken with different roll angles and sometimes
different pointing centers, the exposure
map may show large differences at very large off-axis angles. To avoid
regions of too low exposure,
we only considered regions where the exposure time was larger than 50\% 
of the central value. Furthermore, we truncated the radial profile when
one-third of
the circumference was outside the 50\% exposure region defined above.
As in paper I, we take the position of the BCG closest to the X-ray 
peak as cluster center. 

The new part of the data analysis is described below.

\subsection{Spectral normalization}

In this section we derive the cluster spectral normalization $\eta$ needed to convert
the surface brightness profile into an emissivity profile. The normalization 
weakly depends on $T$, and therefore we perform a spectral fit leaving
$T$ free and we marginalize over $T$.

To measure the spectral normalization, we extracted photons in [0.3-7] 
keV band in two regions: $0.15 < r/r_{500}<0.5$ for the source 
and the annulus with
$r/r_{500}>0.7$ and within the 50\% exposure region for the background. The cluster
central emission was excised to avoid the effects of the possible presence of 
a central cool core, and the outer radius $r/r_{500}=0.5$ was chosen to maximize the
signal-to-noise ratio, following previous literature works (e.g., Croston et al.
2008). The X-ray emission at $r/r_{500}>0.7$ is usually
very low compared to the background. Moreover, by considering a range of
radii, our background estimation is minimally contaminated by the cluster
flux. Croston et al. (2008) and Sun et al. (2009), with which we share
data for three clusters, also used the local
background in their analysis.

The source spectrum was fit with an absorbed APEC (Smith et al. 2001)
plasma model, with the absorbing column fixed at the Galactic value
(Dickey \& Lockman 1990),   
the metal abundance fixed at $0.3$ relative to solar,
and the redshift of the plasma model fixed at optical redshift. 
The fit accounts
for variations in exposure time, excised regions, etc., as mentioned.
The spectrum was grouped to contain a minimum of five counts per
bin and the source and background data were fitted within the 
XSPEC spectral package using the
modified C-statistic (also called W-statistic in XSPEC).
Simulations in Willis et al. (2005)  have
confirmed that this methodology is reliable (and is now used routinely).

One cluster (CL2081) has data that are too poor to derive a spectrum
of quality sufficient to measure a core-excised normalization and therefore
was dropped from the sample.
The impact of the removal of CL2081 (and of one more cluster dropped
in the next section) on our results is
discussed in Sec.~4.4.

\subsection{Gas density profiles and mass}

To measure the gas density profile we assumed a sum of two   
$\beta$ models to allow it to deviate
from a perfect single $\beta$ model (e.g., because of the presence of
a central cool core). Mathematically, the
three--dimensional
profile is given by the sum of two $\beta$ models (with the same slope $\beta$)
\begin{equation}
n^2_e(r) = n^2_{e,0} (1+(r/r_{c,0})^2)^{-3\beta} + n^2_{e,1} (1+(r/r_{c,1})^2)^{-3\beta} \ ,
\end{equation}
where $n_{e,i},r_{c,i},\beta$ parameters have the standard
meaning. We projected the 
three--dimensional profile assuming spherical symmetry 
to obtain the projected gas density. We then converted it into a count-rate
profile using the normalization $\eta$ derived above, and we fit 
it to unbinned X-ray data to derive best fit parameters and uncertainties.
We explored the parameter space by Marcov Chain Monte Carlo.
We preferred projecting the model instead of de-projecting the data because
the latter are noisy and because of the well--known advantages
of projection over deprojection (e.g., Pizzolato et al. 2003; Olamaie et al. 
2015).  Our method builds upon
Mohr et al. (1999) and Ettori et al. (2004) and improves upon these works
by removing simplifying assumptions
(e.g., absence of holes and gaps in the spectral extraction region, a single beta
model, binning, Gaussian errors, etc.). Mathematical details are given in Appendix A.

In practice, we extracted the
distance from the cluster center $r$ of individual
photons in the [0.5-2] keV band, and we fit the projected radial profile without any
radial binning (for the latter see Andreon et al. 2008, 2011, 2016, CIAO-Sherpa, 
and current approaches, such as BayesX, Olamaie et al. 2015, Mantz et al. 2016a). 
Our fit accounts for vignetting, excised regions,
background level, variation in exposure time, 
and Poisson fluctuations using
the likelihood function in Andreon et al. (2008). 

The deprojected 
core-excised gas masses,
$M_{gas,500,ce}=M_{gas}(0.15<r/r_{500}<1)$ are listed in Table~1.
The smallest $0.15r_{500}$ radius is 
49 arcsec, which is much larger than the worse PSF of the instruments used ($\lesssim 8$ arcsec),
so that the contribution of any cool-core flux spilled out 
of the excised region is negligible. Errors are derived
by marginalizing over all model (spatial and spectral) parameters. 

Although our radial profile model is very flexible (four degree of freedom) 
in describing the surface brightness profile of clusters, it assumes a unimodal
X-ray emission and, therefore, cannot deal with
the bimodal CL1022 cluster (Paper I), which we dropped from the sample.

\subsection{Gas mass computation checks}

We extensively tested our gas masses against our derivations
from different X-ray data and  by other authors using the same
or different
X-ray data; we found 
good agreement, as detailed below. Several of the works we used in the comparison 
have made different assumptions on temperature profile, have adopted
a different annulus to compute the brightness-density
conversion, have performed a non-parametric deprojection, have used
a different band for computing the brightness profile, or have binned
the data. The agreement found indicates that differences in assumptions
lead to minor differences to the final result. In particular,

1) Three clusters were observed with Swift and Chandra/XMM. Gas
masses derived by different X-ray data differ by less than $1\sigma$.

2) Wang et al. (2017, in preparation) kindly recomputed the gas mass of CL3046 using, 
as we did, Chandra data from the archive, modifying their pipeline
to adopt the same $r_{500}$ and the same annulus we used to convert 
surface brightness to gas density. The final numbers are identical to ours.

3) The agreement with the unmodified pipeline of Wang et al. (2017, in preparation) 
for the three clusters in common
is better than 0.03 dex for gas masses within their $r_{500}$.
We reproduced the  Croston et al. (2008) XMM-Newton results for CL1047 and CL1052
(at their $r_{500}$) to better than 0.1 dex,  using 
the same XMM data or independent Swift data (CL1047 only). 
Sun et al. (2009) computed the gas mass fraction within $r_{2500}$ of Abell 1238 
and we agree (at their $r_{2500}$) at better than 0.1 dex (however, we 
must point out that we pushed our analysis
beyond its limits, at the cluster center, excised in our analysis in Sec 4).

4) We also compared (Fig.~2) the deprojected 
electron density profiles of two clusters, CL1047 and CL1052, derived by 
Croston et al. (2008, points with error bars) and by us (solid curve
with 68 \% errors shaded). We compared the same XMM-Newton data 
(central and right panel) or data from different telescopes (left panel).
The profiles closely match in all cases. 
The radial profiles
show different degrees of smoothness and errors, the amount of which 
is regulated by the prior, namely the regularization kernel adopted
in Croston et al. (2008)
and by the assumption of a sum of beta models in our work. Therefore,
smoothness and errors can be different for the same analyzed data.

To summarize, the agreement across instruments and with previous works 
makes us confident that our gas mass computation is
solid.

\begin{table}
\caption{Core-excised gas masses.}
\begin{tabular}{l r r l}
\hline
Id & $\log M_{gas,500,ce}$ & err & Telescope  \\ 
 & $[M_\odot]$ & dex & \\
\hline
CL2015  & 12.95 & 0.14 & Swift \\ 
CL2045  & 12.87 & 0.09 & Swift \\ 
CL2010  & 13.09 & 0.08 & Swift \\ 
CL2007  & 12.90 & 0.27 & Swift \\ 
CL3023  & 12.86 & 0.11 & Swift \\ 
CL3030  & 13.61 & 0.03 & Swift \\ 
CL3009  & 12.70 & 0.27 & Swift \\ 
CL1209  & 12.53 & 0.07 & XMM \\ 
CL3053  & 12.40 & 0.15 & Swift \\ 
CL3000  & 13.32 & 0.07 & Swift \\ 
CL3046  & 13.87 & 0.03 & Swift \\ 
        & 13.84 & 0.02 & Chandra \\ 
CL1033  & 12.83 & 0.11 & Swift \\ 
CL1073  & 13.02 & 0.05 & Chandra \\ 
        & 12.97 & 0.10 & Swift \\ 
CL3013  & 13.43 & 0.04 & Swift \\ 
CL1014  & 13.22 & 0.08 & Swift \\ 
CL1020  & 13.15 & 0.06 & Swift \\ 
CL1038  & 13.15 & 0.07 & Swift \\ 
CL1015  & 13.08 & 0.03 & Swift \\ 
CL1120  & 12.60 & 0.21 & Swift \\ 
CL1041  & 13.60 & 0.01 & Chandra \\ 
CL1132  & 12.29 & 0.44 & Swift \\ 
CL1052  & 13.21 & 0.02 & XMM \\ 
CL1009  & 12.98 & 0.03 & Chandra \\ 
CL3049  & 12.70 & 0.08 & Swift \\ 
CL1030  & 12.71 & 0.19 & Swift \\ 
CL1001  & 13.39 & 0.02 & Swift \\ 
CL1067  & 12.98 & 0.09 & Swift \\ 
CL1018  & 12.57 & 0.10 & Swift \\ 
CL1011  & 12.88 & 0.15 & Swift \\ 
CL1039  & 13.25 & 0.05 & XMM \\ 
CL1047  & 13.22 & 0.06 & Swift \\ 
        & 13.15 & 0.02 & XMM \\ 
CL3020  & 13.02 & 0.08 & Swift \\ 
\hline      
\end{tabular} 
\hfill\break
Coordinates, redshifts, masses, and X-ray luminosities are listed
in Paper I.
\end{table}

\subsection{The effect of assuming a $\beta$ value}

Several works (Vikhlinin et al. 2006, Sun et al. 2009, Eckert et al. 2012, Morandi et al.
2015) agreed on a steepening of the gas mass profile close to $r_{500}$, 
although the precise location and amplitude of the steepening differ across studies: 
close to $r_{500}$,  
Vikhlinin et al. (2006) found an effective $\beta=0.78$ and Morandi et al. (2015) found
$\beta=0.67$ (also adopted in our analysis), 
whereas Eckert et al. (2012) reported a value in
between. 
To quantify the impact of the $\beta$ assumed to derive gas masses,
we recomputed them adopting the extreme value $\beta=0.78$. These are smaller
by 0.03 dex with a scatter of 0.01 dex, both of which
are negligible compared to other sources of errors. 
Gas masses depend little on the assumed slope because the 
density profile is constrained by the data and a change in slope  
is compensated by the change of the other four free parameters describing the shape of 
the profile.

\begin{figure}
\centerline{
\includegraphics[width=7truecm]{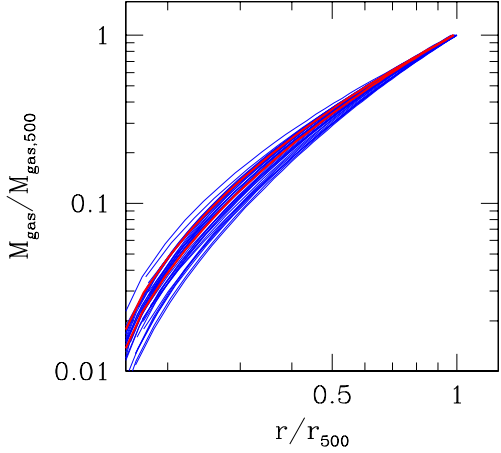}
}
\caption[h]{Integrated gas mass profiles scaled to the core-excised gas mass within $r_{500}$.
The gas mass integration starts at $0.15 r_{500}$. The two clusters with widely different
X-ray brightnesses, yet similarities in mass-related properties of Paper I, are plotted in red.
}
\end{figure}

\begin{figure}
\centerline{
\includegraphics[width=8truecm]{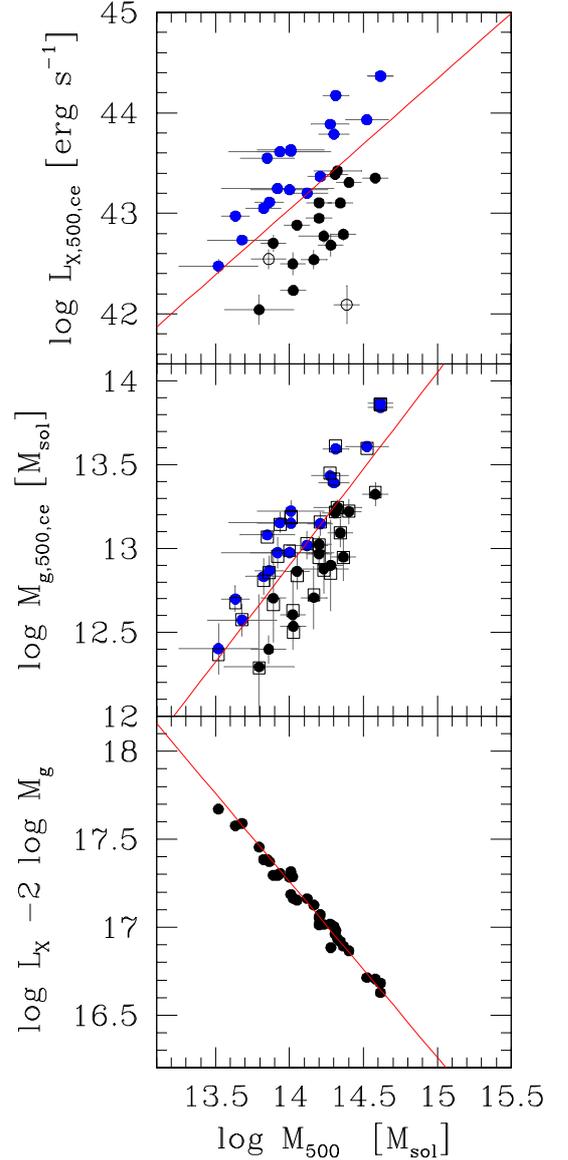}
}
\caption[h]{Core-excised [0.5-2] keV luminosity (upper panel), 
core-excised gas mass (central panel), or a combination of them (bottom panel) vs. caustic mass. 
The red lines indicate the mean fit from this work (bottom panel)
or (top panel) that derived using eq.~4 from the
gas mass slope in Andreon (2010). The
squares in the central panel show the expected gas mass derived
from $L_X$ values in the top panel. 
These are not plotted for the two open points in the top panel,
which are clusters without gas mass
estimates.
Points indicated in blue are above average in $L_X|M$ also turn out to be
above average in $M_{gas}|M$.
Clusters with measurements coming from two X-ray telescopes are shown 
twice, but the two points may be almost perfectly overlapping in one
of the panels (for example, the brightest cluster in the top panel)
and therefore do not show up as such.
}
\end{figure}

\section{Results}

In Sec 4.1 we use our X-ray unbiased sample and 
extend 
the evidence that gas density profiles are similar across 
clusters to the whole class of galaxy clusters, extending previous
claim restricted to subpopulations of clusters.  We use this result to
show that this similarity 
induces a tight covariance
between gas mass (or gas mass fraction) and X-ray luminosity, clarifying
the profound effect of sample selection on the halo gas mass fraction (sect. 4.2).
In Sec 4.3 we compute the mean gas mass fraction and its scatter, unveiling a
more variegate cluster population than seen so far. 

\subsection{Similarities of integrated gas mass profiles}

Clusters may show a large variety of 
X-ray luminosities at a given mass because of differences in a)
the overall gas mass fraction or b) the shape of the 
radial profile (see, e.g., Arnaud \& Evrard 1999).
Fig.~3 removes normalization differences by showing 
the integrated gas mass profiles rescaled 
by $r_{500}$ (in abscissa) and by $M_{gas,500,ce}$ (in ordinate). Profiles
are similar, although not identical, indicating a
strong similarity of cluster gas integrated
density profiles. 
In particular, we discussed in Paper I two clusters, CL2007 and CL3013, that are
significantly different in their X-ray
properties (factor 16 in X-ray luminosity and surface brightness) even though they have
the same mass, velocity dispersion, richness, redshift, X-ray observations of the same depth 
and are located in directions
characterized by the same Galactic absorption. We highlight these clusters in Fig.~3 (red color). 
Once scaled to the total gas mass their profiles  are
entirely consistent with all other profiles in the figure.

The similarity of the scaled gas mass profiles was already 
established for subpopulations of galaxy clusters 
(e.g., X-ray selected, relaxed, etc., Croston et al. 2008; Pratt et al. 2009; 
Mantz et al. 2016b); this evidence is shown here, perhaps for the first time, for
the whole population of clusters at a given mass. 
We emphasize that although we fixed the slope $\beta$ to 
2/3, this only sets the slope of the integrated gas mass profile at large radii 
but the shape of the profile remains flexible because we use four free parameters
(two core radii and two normalizations) to describe the shape. We also emphasize that variations
on small spatial scales are washed out by our analysis; studying these variations would
require a different analysis and data. Furthermore, exceptions
certainly exist, for example, the bimodal CL1022 cluster has 
a different radial profile by definition, since 
it is characterized by two centers.

The similarity of the profiles integrated over the same radial range
generates
a tight, almost scatterless relation between core-excised gas masses
$M_{gas,500,ce}$, core-excised [0.5-2] keV luminosities $L_{X,ce}$, and 
halo masses $M_{500}$, shown in the central panel of Fig.~4,
\begin{equation}
\log L_{X,ce} -2 \log M_{gas,500,ce} +\log M_{500} = 31.26\pm 0.04   \ ,
\end{equation}
where the quoted numbers are the median and semi-interquartile range measured
on our sample\footnote{X-ray luminosity is projected, 
while gas mass is deprojected.}. This relation differs from the well--know
trend between X-ray luminosity and gas mass, where massive clusters tend to
be bright and rich of gas compared to their lower mass counterparts. 
Here, the trend is not driven by mass, the covariance is
at a fixed mass; a cluster that is richer in gas by $\Delta$ than
the average at a given mass (Fig.~4, middle panel) is  
brighter by $2\Delta$ in $L_X$ than the average at a given mass (Fig.~4,
top panel).

A relation (with unknown scatter) between gas mass and X-ray luminosity at a fixed
mass is expected at least since the work of Arnaud \& Evrard (1999): 
the X-ray luminosity can be written as
$L_X \propto f^2_{gas,500} M_{500}$,   
or equivalently 
$L_X \propto M^2_{gas,500} M^{-1}_{500}$,    
which is eq.~2, with a proportionality constant that depends on the cluster structure
($=\langle\rho^2\rangle/\langle\rho\rangle^2$)
and, weakly, on temperature. Since eq.~2 has a small scatter, we
should also see almost the same small scatter for
samples that are not representative of the whole cluster population at a given
mass (X-ray selected, selected to be relaxed, or just
a collection of clusters without any selection function).
In fact, by re-analyzing values published in literature we found
small scatters (0.03 to 0.04 dex) for a)
REXCESS (Pratt et al. 2009; Croston et al. 2008), see also Pratt
et al. (2009) analysis; b) the brightest cluster sample 
of Mantz et al. (2010); and c) clusters in Maughan et al. (2008, see Andreon et al.
2016 for the scatter computation). 
While the scatter in these
samples is small and comparable to what we found, the value derived for XUCS refers 
to the broader population of all
clusters at a given mass. 
In these works, gas masses and halo
masses have been derived in different ways under different
assumptions and, therefore, the small scatter we found is not
due to a hypothetical (and still unidentified) problem in our data or analysis
but is instead the genuine property of clusters. 

The low scatter eq.~2 implies that gas mass can be accurately predicted
from the core-excised luminosity and mass, which is listed in eq. 2 and
enters in the determination of $r_{500}$, as shown in the central panel of Fig.~4 for 
the XUCS. As obvious from the low scatter of Equation~2, the 
gas mass predicted from $L_{X,ce}$ (open squares) is almost coincident with the 
measured gas mass (solid dots).

The low scatter of Equation~2 has two additional implications.
First, 
the residuals in $L_X|M$ must be tightly correlated with the residuals in $M_{gas}|M$
to keep the scatter of eq.~2 low.
In particular, the scatter in $L_X|M$
should be almost exactly twice the scatter in $M_{gas}|M$ (exactly so
if eq.~2 were scatterless). The covariance matrix of $\log L_X - \log M_{gas}$ 
is therefore
\begin{equation} 
  \Sigma = \left(\begin{array}{cc}
  \sigma^2_{L_X} & \rho \sigma_{L_X}\sigma_{M_{gas}} \\
  \rho \sigma_{L_X}\sigma_{M_{gas}} & \sigma^2_{M_{gas}} 
    \end{array}\right)
    \approx \left(\begin{array}{cc}
    4 & 2 \\
    2 & 1 
    \end{array}\right)
    \sigma^2_{M_{gas}} \ , 
\end{equation}
where the second equality is 
because $\sigma_{M_{gas}}\sim 2 \sigma_{L_X}$ and because the two
scatters are almost perfectly correlated
to keep the scatter of eq.~2 low, i.e., $\rho$ is close to $1$. 
The numerical value of $\sigma_{M_{gas}}$ is derived Sec.~4.3.

Second, 
the slopes of the relations between X-ray luminosity and gas mass with halo mass, 
$\beta_{L_X}$ and $\beta_{M_{gas}}$, 
are related by 
\begin{equation}
\beta_{L_X}-1 = 2 (\beta_{M_{gas}}-1) \ 
\end{equation}
because a different relation would induce a scatter in
eq.~2.
This relation requires, however, confirmation by a sample
displaying a larger mass range than probed by XUCS.
 
Coefficients of covariance matrix and slopes in
Mantz et al. (2016b) are consistent with eq.~3 and 4, 
although their analysis is based on a part of the cluster population
only (the subsample of
relaxed clusters) and ignores the necessary accounting of the
effect of sample selection on these quantities. 
We note that $\rho=0$ assumption in Mantz et al. (2016a) disagree
with our results.

\subsection{Bias of the X-ray selection on halo gas mass fraction}

The effect of the selection function on the halo gas mass fraction is
considered unclear in previous works
(Sun et al. 2012, Laudry et al. 2013, Eckert et al. 2016), assumed to be
absent in some scaling relations and cosmological analyses
(e.g., Mantz et al. 2010, 2016a\footnote{Mantz et al. (2016a) account for the
effects of the X-ray selection function on X-ray luminosity but not on gas mass.}), 
or was claimed to be important (e.g., Mantz et al. 2016b, unfortunately without
accounting for it).

Eq.~3 mathematically quantifies that the 
selection function has direct consequences on the halo gas mass fraction;
samples where $L_X$-bright clusters are over-represented are also overly rich in
gas-rich clusters. This is illustrated in  
Figure~4, where clusters have been color-coded 
according to their $L_X|M$. Clusters above average in $L_X|M$ (top panel)
are also above average in their relative gas content for their mass (central panel)
and display a smaller scatter than the whole population.
This is a consequence of the similarity of gas profiles (Figure~3), or, 
equivalently, of the existence of a tight covariance (i.e., $\rho=1$)  between 
$L_X$ and $M_{gas}$. Therefore,
accounting for the X-ray selection is essential to derive
an unbiased halo gas mass fraction and scatter. 
If only the blue points are selected and the effects of the selection function 
are ignored, the mean relation derived would have a higher intercept and a 
significantly smaller scatter than the whole sample. 
Therefore, not accounting for the selection on the halo gas mass fraction
leads to higher average gas masses and a smaller scatter than  
the whole population.  
It is also evident that recovering the red
line in the bottom panel, and the full width of the vertical distribution
at a given mass using only the blue points would be hard because 
it would require reconstructing the full range and the correct mean
from just the tail of the distribution, while relying heavily on
strong assumptions of the properties of the unobserved part of the 
population.

Therefore, an X-ray selection that is not
properly accounted for can have profound implication on the resulting average gas
fraction. Moreover, we can also dismiss the 
qualitative argument that the X-ray selection is negligible because
most of the gas mass comes from faint X-ray regions 
and these are not triggering the X-ray detection.
It can be easily proven that all radii outside
the core carry similar contributions to the gas mass. 
Moreover, since the shape of the normalized radial profiles are
similar, the gas mass is equally constrained by the X-ray brightness 
at any radius, not only by those where the X-ray emission is faint.

\begin{figure}
\centerline{
\includegraphics[width=7truecm]{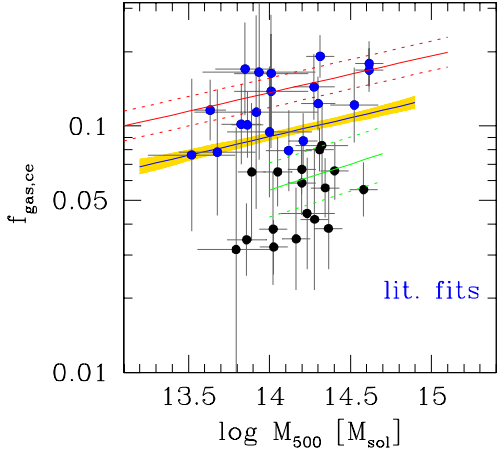}
}
\caption[h]{
Core-excised gas mass fraction $f_{gas,ce}$ vs. halo mass $M_{500}$ of XUCS
sample with fits to other samples.
The black line with shading (error) represents the Andreon (2010) fit to clusters 
in Vikhlinin et al. (2006) and Sun et al. (2009), the red line represents the 
Landry et al. (2012, dashed lines indicate scatter) fit to their sample,  and
the green line represents the 
XXL 100 brightest clusters (Eckert et al. 2016, dashed lines indicate uncertainty); 
see text for details. Blue points are clusters that are brighter in X-ray than
the average for their mass.
}
\end{figure}

Figure~5 illustrates the various results one may obtain with different accounting
of the selection function on the halo gas mass fraction. The points are 
core-excised gas mass fractions of the XUCS clusters. 
This sample, because it is X-ray unbiased, is not affected by
the X-ray selection. The clusters show a large scatter in gas mass fraction
at a given mass that we quantify in the next section. If only clusters
above average in $L_X|M$ were available (blue points), the scatter
would be biased down and the mean gas mass biased high.
In the same figure we plot best fits and error regions obtained by different 
authors from samples with different selection criteria, but similar mass range.
Landry et al. (2013) obtained the 
best fit at the top of the diagram by studying 
the most luminous clusters and neglecting the selection function. 
The representative XUCS sample, which has systematically lower gas mass fractions,
confirms the selection bias of the Landry et al. (2013) sample; this possibility
is also mentioned, along with others, by the authors.
The bottom (green) solid lines in Fig.~5 show the fit to the 100 brightest
clusters in the XXL survey (Eckert et al. 2016). 
Their mean relation is significantly lower, and it appears to be even 
below the average of our X-ray unbiased sample, which is surprising because
the X-ray selection of their sample should favor clusters with large
gas mass fractions. 
This may happen if their weak-lensing cluster masses 
are overestimated; this possibility
is also mentioned, along with others, by the authors. 
Finally, we also plot the fit derived by Andreon (2010)
for the sample in Vikhlinin et al. (2006) and Sun et al. (2009). This sample
is formed by relaxed cluster but, apart from that, has an unknown selection
function. As quantitatively discussed in next section, our sample (with
available selection function) returns a mean gas mass in close
agreement with theirs. The fits on samples by
Landry et al. (2013), Eckert et al. (2016), Vikhlinin et al. (2006), and Sun et al. (2009) 
are based on total gas masses, while we plot 
core-excised gas masses for XUCS. However, this mismatch does not impact our results
because of the negligible differences between them for our sample. In fact, 
including the central emission under the assumption of a constant temperature 
(which overestimates the gas mass, because excess emission is more
likely due to a decrease in temperature rather than higher gas density)
would result in 
gas masses that are higher by $\sim 0.012$ dex,
with a small scatter of $0.003$ dex (semi-interquartile range).

To summarize, the sample selection has the same effect on gas mass than on X-ray luminosity. 
Clusters for which it is easier to collect large numbers of photons
have on average larger-than-average gas masses; if not
accounted for, gas mass is biased high and the scatter biased low. 
The amplitude of the bias is hard to infer from an X-ray selected sample because 
the width of the $L_X|M$ or $M_{gas}|M$ distribution needs to be either reconstructed 
from its tail or relies heavily on assumptions and expectations about 
the behavior of the unobserved population. 
A similar situations occurs for studied quantities showing a covariance
with the quantity used for selecting the sample, such as X-ray luminosity or the
strength of the SZ signal when studying the scaling with gas mass.  If the sample is
X-ray unbiased, i.e., clusters that are brighter-than-average and fainter-than-average
for their mass enter with the same probability, the sample selection
does not affect the derivation of the halo gas mass fraction or scatter.

\subsection{Gas mass fraction: Mean and scatter}

\begin{figure}
\centerline{
\includegraphics[width=7truecm]{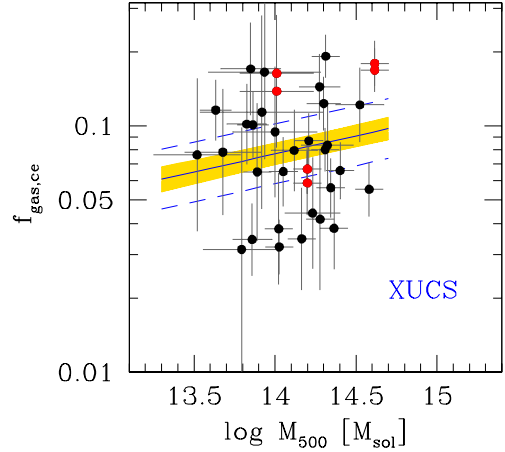}
}
\caption[h]{Core-excised gas mass fraction $f_{gas,ce}$ vs. halo mass $M_{500}$ of XUCS
sample. 
The solid line indicates the mean relations, the shading indicates its 68\% uncertainty,
whereas the dashed lines indicate
the mean relation $\pm 1\sigma_{intr}$. 
Close pairs of red points (with identical masses) indicate measurements
from different X-ray telescopes.
}
\end{figure}

In this section we fit the relation between 
gas mass fraction and mass (points in Fig.s~5 and 6), allowing 
intrinsic scatter in $f_{gas}|M$. We use a 
linear model with intrinsic scatter $\sigma_{intr}$ that
also allows for measurement errors (Dellaportas \& Stephens, 1995), 
which were already used in Andreon (2010; where we also distributed the fitting code)  
to fit values for clusters in Vikhlinin et al. (2006) and Sun et al. (2009). We 
fit the data in the gas mass versus halo mass plane, where errors are less
correlated (see Andreon 2010). 
For intrinsic scatter, and 
$\log$ of gas mass fraction at $\log M/M_\odot=14$, $\log f_{gas,14}=\log M_{gas}-14$, 
we assume a uniform and wide range of values that certainly 
includes the true value. 
Since we cannot properly determine the slope of the relation in the limited range
covered by XUCS clusters, we adopt as prior the posterior derived in 
Andreon (2010) for the sample in 
Vikhlinin et al. (2006) and Sun et al. (2009), $0.15\pm0.03$.
As shown in Sec.~4.4, the assumption of the slope value does not
change the results because the clusters studied have similar masses.
The parameter space is sampled by Gibbs sampling using JAGS
(see Andreon 2011). For the three clusters for which multiple  
estimates for gas mass fraction are available, which are derived
from different telescopes, the fit only uses those with smaller errors.
Our errors on $\log M^{obs}$ incorporate both statistical
error and a 20\% intrinsic scatter between caustic masses and true masses.

We found
\begin{equation}
\log f_{gas} =  (0.15\pm0.03)\  (\log M_{500}-14) -1.10\pm0.04 \ ,
\end{equation}
with $\sigma_{intr}=0.17\pm0.04$, signaling that the slope posterior is
largely determined by the slope prior. 

The results of the fit are plotted in Fig.~6, including 
the mean relation (solid
line), its 68\% uncertainty (shading), and the mean
relation $\pm 1 \sigma_{intr}$. 

There is good agreement between the {\it mean} scaling derived for the X-ray
unbiased sample and for clusters in Vikhlinin et al. (2006) 
and Sun et al. (2009) samples derived in Andreon (2010), which are
reproduced on the top of 
XUCS clusters in Fig.~5 (the same data shown in Fig.~6). The values of $\log$ of gas mass fraction at 
$\log M/M_\odot=14$ of the two samples
differ by $-0.05\pm0.05$ dex, suggesting that 
the average gas mass fraction is similar in the two samples and that 
the sample in Vikhlinin et al. (2006)
and Sun et al. (2009) does not systematically select gas-rich or gas-poor
clusters. The consequences that this agreement has on the hydrostatic mass bias 
is presented in
the companion paper (Andreon et al. 2017). There is also agreement on the slopes, but
this occurs because the slope is unconstrained by XUCS data and we assumed 
the posterior derived for the sample in 
Vikhlinin et al. (2006) and Sun et al. (2009).

The scatter is robustly measured because our sample selection function does not
favor gas-rich clusters at a given mass and, in addition, the
selection function is easily characterized at a given mass (a constant).
Therefore, we do not need to make assumptions about the unseen population as
in X-ray selected samples and
the intrinsic scatter is, by large, simply the vertical distribution
of the data points after proper accounting of the errors. 
Our estimate of the scatter, $0.17\pm0.04$ dex, 
indicates that there is a variable, cluster-to-cluster
amount of gas within $r_{500}$.  
The scatter we found on the gas mass fraction is larger than values
derived in literature, which however only refers to   
subsamples of the whole cluster population, for example 
relaxed clusters (e.g., Vikhlinin et al. 2006, Sun et al. 2009, 
Andreon 2010, Mantz et al. 2016b). For example,  
clusters in Vikhlinin et al. (2006) and Sun et al. (2009) show
a gas mass scatter of $0.06\pm0.01$ dex (Andreon 2010).
Our X-ray unbiased sample thus unveils a larger variety in intracluster
medium content than past samples focused on subsamples.
The large scatter in gas mass fraction could be interpreted 
as a different fraction of gas being expelled beyond the $r_{500}$, by, for example,
a different amount of 
feedback from AGN heating the IGM to temperatures in the $10^8-10^{8.5}$
range (Le Brun et al. 2014).
Alternatively, gas mass fraction
differences are present ab initio in the volume occupied by the cluster,
but there is no evidence for that in simulations.

The pseudo pressure parameter, $Y_X=M_{gas} T$, is believed to be a low-scatter
mass proxy (Kravtsov et al. 2006). In Kravtsov et al. (2006) simulations $Y_X$
has a $0.03$ dex scatter with mass, even lower than gas mass scatter, $0.06$ dex, because
gas mass and temperature are anti-covariant in these simulations.
Our gas masses display a scatter that is much larger than Kravtsov et al. (2006) simulations, and 
simulations  by other authors (e.g., Stanek et al. 2010) do not show the anti-covariance between
$T$ and $M_{gas}$ that reduces the scatter of $Y_X$ below the gas mass scatter, 
as in Kravtsov et al. (2006).
Furthermore, more recent simulations (e.g., Le Brun et al. 2014)
suggest, in agreement with our data, that 
$Y_X$ is affected by the ICM physics.
Therefore, the large gas fraction scatter we found, together with the uncertain theoretical
covariance between gas mass and temperature, makes $Y_X$ a questionable
low-scatter mass proxy. The large gas fraction scatter we found also similarly badly
affect SZ proxies, such as $Y_{SZ}=M_{gas} T_m$ (Nagai 2006).

As already mentioned, in scaling relations using X-ray selected samples,
or cosmological analyses based on them,  
the width and center of the distribution (at a fixed mass) 
of the observable used to select
the sample cannot be inferred from the
observation of just a tail, i.e., 
the degeneracy between intrinsic scatter and intercept cannot be broken. 
A way to break the degeneracy, i.e., to derive intercept {\it and} scatter,
is to have an estimate of the size of the whole population (assuming
a cosmology, for example) and a assumption about the shape of
the scatter (lognormal, for example). The latter assumption is strong 
when the whole population (at a given mass) is not entirely seen, but weak when, as
in our case, the whole population is observed. Alternatively, one may
assume a prior from simulations, such as 
the Vikhlinin et al. (2009) cosmological analysis. 
Our measurement of the intrinsic scatter posterior offers
a valid alternative to the use of simulation values, which, as mentioned
above, also depend on the implemented subgrid physics. 

As mentioned in the introduction, in Paper I we found an amazing variety of
X-ray luminosities at a given mass. The similarity of the gas mass density
profiles (sec.~4.1) indicates that the observed large variance is not
due to the way the gas mass is distributed within $r_{500}$. 
Differences in the overall gas mass fraction are mainly
responsible for the wide range 
of core-excised X-ray luminosities observed in our previous work
because the predicted (from eq.~3) scatter in $L_X|M$ is
expected to be twice the gas mass scatter, i.e., $0.34\pm0.08$
(versus the measured $0.47\pm0.07$ dex). To summarize, overall gas
differences, from cluster to cluster, are the main source of the variance
in core-excised luminosities seen in Paper I.

\subsection{Sensitivity analysis}

In our analysis, we started from a sample that was not selected based on its
X-ray properties. However, we later discarded two (out of 34) clusters 
because of their X-ray properties.

However, we verified that this does not alter the original properties of the
sample. In fact, the almost scatter-less relation between
gas mass, X-ray luminosity and halo mass (eq.~2) may, at most,
acquire two outliers; two objects are too few to erase the bias of X-ray
selection on the halo gas mass fraction (sec.~4.1); they cannot 
remove the scatter in the $M_{gas}-M$ relation (sec.~4.3).
We further checked that the intercept of eq.~5 changes by 
less than 0.01 dex if we reintroduce the two clusters, 
using a gas mass predicted from  $L_X$
(sec 4.1).

Our analysis assumes that caustic masses have 20\% scatter, as supported by
numerical simulations and observations (e.g., Serra et al. 2011, Gifford \& Miller 2013,
Geller et al. 2013, Maughan et al. 2016).
In Andreon et al. (2017) we present a more hypothesis-parsimonous analysis where
we leave the scatter free to vary and we marginalize over it. This more rigorous 
approach confirms the 
present result of a large gas fraction scatter and therefore
a high variable, cluster-to-cluster, amount of gas 
within $r_{500}$.

Finally, our analysis of the mean gas mass fraction and scatter assumed
as slope prior the posterior of Andreon (2010)
based on Vikhlinin et al. (2006) 
and Sun et al. (2009) clusters. 
If we instead take a
uniform prior on the angle to allow  different
slopes, we find 
an identical intrinsic scatter, 
a very similar intercept $-1.17\pm0.05$, as well
joint posterior distributions of gas mass fraction and intrinsic scatter
close to those derived in our standard analysis.
This shows the robustness of our 
conclusions on assumptions about the slope.

\section{Conclusions}  

We used a cluster sample formed by 34 clusters observed in X-ray in Paper I  
with caustic masses and whose selection is, at a given
cluster mass, independent of the intracluster medium content. 
We derived gas masses by projecting a flexible radial
profile and fitting its projection to the unbinned X-ray data and propagating
all modeled sources of uncertainties (e.g., spectral normalization, variation in exposure
time including those originated by vignetting or excised regions) 
with their non-Gaussian behavior
(when relevant) into the gas mass estimate using
Bayesian methods.

We found that integrated gas density profiles (i.e., the profile shapes)
are quite similar, although not identical, extending
results based on samples that do not explor the full range of the cluster
population at a given mass (e.g., relaxed-only, or X-ray selected, 
clusters). The overall gas mass fraction is, instead, quite different from cluster
to cluster, because the scatter is $0.17\pm0.04$ dex. 
The scatter we found is larger than found
in published samples not exploring the full range of the cluster
population at a given mass and 
indicates a quite variable amount of feedback from cluster to cluster.
Our X-ray unbiased
cluster sample thus unveiled a part of the cluster population that is
underrepresented, when not missed altogether, in X-ray selected samples.
 
The similarity of the integrated gas density profiles implies a tight relation between
core-excised X-ray luminosity, gas mass, and halo mass (eq.~2).
This implies a) a specific form for the  $\log L_{X,ce}$ $\log M_{gas,500,ce}$
covariance matrix in which the scatter in gas mass is twice the scatter in X-ray
luminosity and the correlation coefficient $\rho=1$; b) that 
the slopes of X-ray luminosity and gas mass with halo mass, 
$\beta_{L_X}$ and $\beta_{M_{gas}}$, 
are related by a simple expression (eq.~4);
and c) that the effect of the selection function on the halo gas mass fraction is the
very same as for X-ray luminosity, i.e., gas-rich clusters are preferentially
included in X-ray selected samples.

Finally, the overall gas mass fraction scatter is the main cause of the
large variety of core-excised X-ray luminosities
observed in our previous work, rather than the way in which
the gas mass is distributed within $r_{500}$.

\begin{acknowledgements}

SA dedicates this work to the memory of his father Duilio. We thank Adam Mantz for
discussion on his papers.

\end{acknowledgements}

{}

\appendix

\section{Calculating gas densities profiles and masses}

Let us consider a  three-dimensional
profile given by the sum of two $\beta$ models (with the same slope $\beta$)
\begin{equation}
n^2_e(r) = n^2_{e,0} (1+(r/r_{c,0})^2)^{-3\beta} + n^2_{e,1} (1+(r/r_{c,1})^2)^{-3\beta} \ ,
\end{equation}
where $n_{e,0}, n_{e,1}$ are the central densities and $r_{c,0},r_{c,1}$ are the
core radii.

The projection in two dimensions is analytically known 
\begin{equation}
\Sigma(b) = \Sigma_0 (1+(b/r_{c,0})^2)^{-3\beta+0.5} + \Sigma_1 (1+(b/r_{c,1})^2)^{-3\beta+0.5} +
\Sigma_{bkg} \ ,  
\end{equation}
where the last term accounts for the background\footnote{For numerical reasons and to improve chain mixing
is preferable to reparametrize the model using 60 and 5 arcsec as pivot radii: 
$\Sigma(b) = \Sigma_0(60) (1+(60/r_{c,0})^2)^{3\beta-0.5} (1+(b/r_{c,0})^2)^{-3\beta+0.5} + 
\Sigma_1(5) (1+(5/r_{c,1})^2)^{3\beta-0.5} (1+(b/r_{c,1})^2)^{-3\beta+0.5} + \Sigma_{bkg} $. 
Inside the MCMC sampler and for numerical computation we use this expression.}.

The relation between the normalizations $n^2_{e,i}$ and $\Sigma_i$ is given by
\begin{equation}
n^2_{e,i} = \frac{\Sigma_i}{r_{c,i} \sqrt{\pi} } \frac{\Gamma{(3\beta)}}{\Gamma{(3\beta-0.5)} } = 
\frac{c_i k }{r_{c,i} \sqrt{\pi} } \frac{\Gamma{(3\beta)}}{\Gamma{(3\beta-0.5)} } = \frac{c_i}{r_{c,i}}k' 
\end{equation}
(Cowie et al. 1987; Hudson et al. 2010; Ecker et al. 2016), where the second equality accounts 
for the fact that  we do not measure projected
densities $\Sigma_i$'s, but brightnesses $c_i=\Sigma_i/k$ whereas the third
equality comes from defining $k'=\frac{k}{\sqrt{\pi}}\frac{\Gamma{(3\beta)}}{\Gamma{(3\beta-0.5)} }$.
The symbol $\Gamma$ is the usual Gamma function.
Therefore, a fit to the radial
profile returns the core radii $r_{c,i}$,  $\beta$ values, and  $n^2_{e,i}$ coefficients,
the latter up to a constant $k'$ that we now determine. 
We derive it from the normalization of the spectral fit $\eta$ 
(returned by XSPEC, as described in section~3.1) defined
by  
\begin{equation}
\int_\Omega n_e n_p dV = 4 \pi d^2_A (1+z)^2 \eta 10^{14} \ , 
\end{equation}
where $d_A$ is the cluster angular distance.
In fact, by definition
\begin{equation}
\int_\Omega n_e n_p dV = 0.82 \int_\Omega n^2_e dV \ , 
\end{equation}
where 
we assume $n_p=0.82 n_e$ and the integral is
over the ``cylinder" having as base the spectral extraction
region $\Omega$ and extends all along the line of sight.

Equating the right-hand sides of eq.s~A.4 and~A.5, inserting eq.~A.1 into the latter, 
and substituting $n^2_{e,i}$ with the right-hand side of eq.~A.3 gives
\begin{eqnarray}
k' &= 
\frac{4 \pi d^2_A (1+z)^2 \eta 10^{14}}{0.82 \int_\Omega [ \frac{c_{0}}{r_{c,0}} (1+(r/r_{c,0})^2)^{-3\beta} + \frac{c_{1}}{r_{c,1}} (1+(r/r_{c,1})^2)^{-3\beta} ] dV} \ , 
\nonumber\\
 &= \frac{4 \pi d^2_A (1+z)^2 \eta 10^{14}}{0.82 c_\eta} \ ,
\end{eqnarray}
where $c_\eta=\int_\Omega [\dots] dV$ is the net counts in the imaging extraction band
in the spectral extraction
region used to compute $\eta$, which is known.

Since there are no left unknowns, deprojected gas masses can now be 
straightforwardly computed from 
\begin{equation}
M_{gas}(<r')= \int^{r'}_0 4 \pi r^2 n_{gas}(r) dr = \int^{r'}_0 4 \pi r^2 \mu_e m_p  n_e(r) dr \ , 
\end{equation}
where $\mu_e m_p$ is the mean atomic mass, starting the integration from 
$0.15r_{500}$ when core-excised masses are needed.

The parameter
$\beta$ is fixed at $2/3$ because our data cannot constrain
all model parameters for every cluster when using the sum of two beta models to
describe the profile. Sec.~3.4 explores the effect of this assumption. For the 
other parameters, we took weak priors, and, in particular, we took a uniform prior
for $\Sigma_i$'s and $\Sigma_{bkg}$, constrained to be positive (to avoid unphysical
values), and a Gaussian prior on the core radius $r_{c,1}$ with the center equal to $r_{500}/5$
(suggested by Ettori et al. 2015), with a 30\% sigma (i.e., $r_{c,1} \in  
(0,r_{500})$ with 99 \% probability). Finally, we ask that the second component
(modeling the cool core) is much smaller than the cluster itself (i.e., $r_{c,0}<r_{c,1}/5$).
A posteriori, all (posterior mean) $r_{c,0}$ and $r_{c,1}$ are well within these ranges.
Chandra data
also require us to excise the inner 3-5 arcsec region because of the
unmodeled BCG emission.

\end{document}